\documentclass[%
 aip,
rsi,%
 amsmath,amssymb,
 reprint,%
]{revtex4-1}
\usepackage{booktabs}
\usepackage{multirow}
\usepackage{graphicx}
\usepackage{color}
\usepackage{setspace}
\usepackage{ragged2e}
\usepackage{soul}
\newcommand{\cpc}{Comp. Phys. Commun.\ }

\AtBeginDocument{
\heavyrulewidth=.08em
\lightrulewidth=.05em
\cmidrulewidth=.03em
\belowrulesep=.65ex
\belowbottomsep=0pt
\aboverulesep=.4ex
\abovetopsep=0pt
\cmidrulesep=\doublerulesep
\cmidrulekern=.5em
\defaultaddspace=.5em
}

\begin{document}
\title{Characteristic features of the Shannon information entropy of dipolar Bose-Einstein condensates}

\author{Thangarasu Sriraman}
\affiliation{Department of Physics, Bharathidasan University, Tiruchirapalli 620024, India.}

\author{Barnali Chakrabarti}
\affiliation{Department of Physics, 86/1 College Street, Presidency University, Kolkata 700073, India.}

\author{Andrea Trombettoni}
\affiliation{CNR-IOM DEMOCRITOS Simulation Center, Via Bonomea 265, I-34136 Trieste, Italy.}
\affiliation{SISSA, Via Bonomea 265, I-34136 Trieste, Italy.}
\affiliation{INFN, Sezione di Trieste, Via Bonomea 265, I-34136 Trieste, Italy.}

\author{Paulsamy Muruganandam}
\affiliation{Department of Physics, Bharathidasan University, Tiruchirapalli 620024, India.}

\begin{abstract}
Calculation of the  Shannon information entropy $(S)$ and its connection with the order-disorder transition, and with inter-particle interaction provide  a challenging research area in the field of quantum information. Experimental progress with cold trapped atoms has corroborated {this} interest. In the present work, $S$ is calculated for the Bose-Einstein condensate (BEC) with dominant dipolar interaction for different dipole strengths, trap aspect ratio and  number of particles ($N$). Trapped dipolar bosons in an anisotropic trap  {provide an example} of system where the effective interaction is strongly determined by the trap geometry. The main {conlcusion}  of the present calculation is {that} the anisotropic trap  reduces the number of degrees of freedom, {resulting in  more ordered configurations}. {The Landsberg's order parameter} exhibits quick saturation with the increase in scattering length in both prolate and oblate traps. We also define the threshold scattering length which makes the system completely {disordered. Unlike non-dipolar BEC in a spherical trap, we do not find a universal linear relation between $S$ and $\ln N$, and we, therefore, introduce a general quintic polynomial fit rather well working for a wide range of particle number.}
\end{abstract}

\maketitle

\section{Introduction}
\label{sec:1} 

The Heisenberg uncertainty relation and its generalization by Robertson 
{provide} the well known uncertainty relation 
related with the precision of measurement. Later Bialynicki-Birula-Mycielski (BBM)\cite{Bialynicki1975} have obtained a generalization of the uncertainty relation in terms of the information entropies of position and momentum space. The information entropy for a continuous probability distribution $\rho(x)$ in one dimension is defined by 
\begin{equation}
S= -\int_{}^{}\rho(x)\ln{\rho(x)} dx,
\end{equation}
where
\begin{equation}
\int \rho(x) dx=1.
\end{equation}
$S$ basically measures the uncertainty of the corresponding probability distribution. The {entropic uncertainty relation (EUR)} is an important step in this direction and it states that for a three dimensional system
\begin{equation}
S_{r}+S_{k}\geqslant 3(1+\ln{\pi})\cong 6.434,
\label{eq3}
\end{equation}
where $S_r$ and $S_k$ are the entropies in position and momentum space, respectively. The corresponding one-body distributions $\rho(x)$ and $n(k)$ are normalized to one. The lower bound in Eq.~(\ref{eq3}) strictly holds for a Gaussian density distribution. The physical meaning of the inequality is the diffuse density distributions in momentum space is associated with localized density distributions in configuration space and vice versa. 
Thus, the maximal value of $S_{r}$ corresponds to a uniform distribution and the minimal $S_{r}$ corresponds to delta-like distributions having the {minimal} uncertainty. 
The interpretation of $S_{k}$ is the same as that of $S_{r}$. 
However, EUR is a strengthened version of Heisenberg's uncertainty principle, 
EUR does not depend on the state of the system.

Information theory finds an important role in expounding various concepts 
in quantum mechanics. In atomic physics, the information-theoretical concepts 
have been used as the tools to synthesize and analyze the electron densities 
in both momentum as well as coordinate spaces\cite{Gadre1979, Sears1981, Koga1983, Parr1986, Guevara2003, Guevara2005, Sagar2005, Sen2005, Ho2000, Sagar2006}. 
Sears and Gadre  applied the maximum entropy principle for constructing the Compton profiles of atoms and molecules constrained to various moments~\cite{Gadre1979, Sears1981}. They found that maximization of Shannon's entropy subjected to the constraints of average momentum and average energy {is itself sufficient} to approximately construct the Compton profiles from the experimental or theoretical data. Further, information entropy using Thomas-Fermi theory, maximization of atomic information entropy in momentum and configuration spaces, and other features of atomic information entropies have been explored\cite{Gadre1984, Gadre1985a, Gadre1985b, Gadre1985c}.
The group of Sagar {performed several works to calculate} 
information entropy, local correlation, and the measure of mutual information
 in atomic system\cite{Guevara2003, Guevara2005, Sagar2005}. 
Characteristic {features of Shannon information entropy of confined atoms have also been discussed in the reference}\cite{Sen2005}.
Using density functional formalism, Sears {\em et al.} have established\cite{Sears1980} the quantum mechanical kinetic energy as a measure of information in a distribution and further that the quantum mechanical variational principle is a principle of minimum information. Based on information theory, Maroulis {\em 
et al.} have proposed a way to evaluate a basis-set quality in terms of various expectation values and provided a method for improving the quality of a wave function\cite{Maroulis1981}. Following this approach, Simas {\em et al.} 
tested the quality of various orbital basis sets {for helium atom}\cite{Simas1983}.

Very recently, attention has been shifted to the calculation of information entropy in various many-body systems. The group of Massen and Panos have made extensive calculations for the fermionic and bosonic systems and had found a link between information entropy and kinetic energy. They have studied a broad class of fermionic systems like a nucleon in a nucleus, a $\Lambda$ particle in hypernucleus and an electron in an atomic cluster and also the correlated boson-atoms in a harmonic trap~\cite{Massen1998, Massen2001, Massen2002}. The universal trend of the information entropy for all the above mentioned many-body systems was reported and a functional dependence of total $ S=S_r+S_k $ was presented. 
{Typically,} for systems with $N$ particles 
the same functional form $S=a+b\ln{N}$ holds, while the two constants 
$a$ and $b$ vary from system to system.

In the present manuscript, we are interested in the study of dipolar 
Bose-Einstein condensate (BEC).  {Studies of Massen and Panos  on BEC 
consider the dilute interacting Bose gas with the $s$-wave contact 
interaction, which is characterized by the inter-atomic scattering length $a$.} 
However, the experimental observation of BECs of $^{52}$Cr, $^{164}$Dy and $^{168}$Er with the large dipolar interaction has been reported and extensively studied\cite{Gries-cr, Lu-dy, Aika-er}. {The properties of 
the dipolar BEC strongly deviates from 
that of a non-dipolar BEC.} The inter-atomic interaction of the dipolar BEC is now characterized by the $s$-wave contact interaction and an anisotropic long-range dipolar interaction. The anisotropic dipolar BEC has many distinct features. The stability of nondipolar BEC is solely determined by $a$. Positive $a$ corresponds to repulsive BEC which is always stable, while negative $a$ corresponds to attractive BEC and experiences collapse. The stability of a dipolar BEC strongly depends on the trap geometry. In a disk-shaped trap, the dipolar BEC is more stable whereas a cigar-shaped trap always yields an attractive interaction and finally leads to collapse. The peculiar competition between the isotropic short-range contact interaction and anisotropic long-range dipolar interaction makes the system very interesting. 

In this paper, we shall calculate the Shannon entropy for dipolar Bose-Einstein condensates for various system parameters namely trap aspect ratio $\lambda$, the number of constituent particles $N$, the inter-atomic scattering length $a$, and characteristic dipole length $a_{dd}$. We compute the 
Landsberg's order parameter $\Omega$ for the case of dipolar BEC for different trapping geometries and also for various values of inter-atomic scattering length $a$. It has been found that depending on the trap geometry, there is a threshold for the inter-atomic scattering length denoted as $a_{thres}$, below which the Landsberg's order parameter becomes close to zero, $\Omega \approx 0$, 
that is, the system exhibits complete disorder. {Dipolar Bose-Einstein condensates do not exhibit the linear relationship between $S$ and $\ln N$ that was reported by Massen and Panos earlier in connection with several many-body systems.}  For the present study of dipolar BEC, we {introduce} a quintic polynomial fit where several entropy measures strongly depend on the trap geometry and the characteristic dipole length $a_{dd}$ defined in next section. Following the earlier work for fermionic many-body systems\cite{Massen2001}, we establish a link between $S$ and the total kinetic energy $T$ for the case of dipolar BECs. We find that our numerical results can be well fitted by a cubic polynomial and we do not observe any universal behavior as observed in atomic clusters. Also, we notice that the $S$ versus $T$ curves show a strong dependence on both the trap geometry as well as the strength of dipolar interaction. 

It is a well-established fact that the entropy uncertainty relation and different applications of Shannon entropy are {good} tools to correlate the complexity of a system with the inter-particle potentials. In the same way, we utilize a similar kind of measure for the dipolar BEC in different trap geometries, where the complexity of the system is not uniquely determined by the inter-atomic interaction alone. Dipolar BEC is a {system} where even the repulsive condensate may become unstable due to the anisotropic dipole-dipole interaction. Thus the main idea is to correlate the different measures of entropy not only with the fundamental inter-atomic potential but also to give some additional insights about the complexity of the systems, which are further addressed through the calculation of order-disorder parameter.

The organization of the paper is as follows: After the introduction, in Sec~\ref{sec:2}, we present a description of the mean-field model equation for dipolar BECs. In Sec.~\ref{sec:3}, we present the calculation of the Shannon information entropy for dipolar BEC of $^{52}$Cr and $^{164}$Dy atoms followed by a discussion on the various results obtained for different system parameters. Finally, in section.~\ref{sec:4}, we present a brief summary and conclusion.


\section{The mean-field Gross-Pitaevskii equation with dipole-dipole interaction}
\label{sec:2}
At zero temperature, the static and dynamic properties of a BEC can be well described by mean-field Gross-Pitaevskii (GP) equation\cite{Dalfovo1999}. At such 
temperatures, the properties of a dipolar Bose-Einstein condensate of $N$ atoms of which each mass $m$, can be modeled by a mean-field GP equation with nonlocal nonlinearity of the form {(see, e.g., Refs.~\onlinecite{Yi2006, Van2007})}:
\begin{align}
\mathrm{i}\hbar\frac{\partial \phi({\mathbf r},t)}{\partial t} & = \left[-\frac{\hbar^2}{2m}\nabla^2+V_{\text{trap}}({\mathbf r}) 
+ \frac{4\pi\hbar^2a N}{m} \vert \phi({\mathbf r},t)\vert^2 \right. \notag \\
 & + \left. N \int U_{\mathrm{dd}}({\mathbf r}-{\mathbf r}') \vert\phi({\mathbf r}',t)\vert^2 d{\mathbf r}' 
\right]\phi({\mathbf r},t),
\label{eqn:dgpe}
\end{align}
where $ \phi({\mathbf r},t)$ is the condensate wave function with the normalization condition $\int d{\bf r}\vert\phi({\mathbf r},t)\vert^2=1$. In Eq.~(\ref{eqn:dgpe}), the trapping potential $V_{\text{trap}}({\mathbf r}) $ is assumed to be 
of the form
\begin{align}
V_{\text{trap}}({\mathbf r}) = \frac{1}{2} m \left(\omega_x^2 x^2+\omega_y^2 y^2+ \omega_z^2 z^2 \right) \notag
\end{align}
where $\omega_x, \omega_y $ and $\omega_z$ are the trap frequencies and $a$ the inter-atomic scattering length.
The dipolar interaction, for magnetic dipoles, is given by \cite{tuning-ddi,Goral2002} 
\begin{align}
U_{\mathrm{dd}}({\bf R}) = \frac{\mu_0 \bar \mu^2}{4\pi}\frac{1-3\cos^2 \theta}{ \vert {\bf R} \vert ^3} 
\left( \frac{3 \cos^2{\varphi}-1}{2}\right),
\end{align}
where ${\bf R= r -r'}$ determines the relative position of dipoles and $\theta$ is the angle between ${\bf R}$ and the direction of polarization, $\mu_0$ is the permeability of free space and $\bar \mu$ is the dipole moment of the condensate atom. The $\varphi$ is the angle between the orientation of dipoles and $z$-axis. We consider the polarization of magnetic dipoles along the direction of $z$-axis as long as $\varphi = 0$. Nevertheless, it is tunable to change the dipolar interaction from attractive to repulsive. 

To compare the contact and dipolar interactions, often it is convenient to introduce the length scale $a_{\mathrm{dd}}\equiv \mu_0 \bar \mu^2 m/(12\pi \hbar^2)$~\cite{Koch2008}. Chromium has a magnetic dipole moment of $\bar \mu = 6\mu_B$ ($\mu_B$ is the Bohr magneton) so that $a_\mathrm{dd} \simeq 16a_0$, where $a_0$ is the Bohr radius. The dipole-dipole interaction strength is expressed as 
\begin{align}
D=3 N a_{\mathrm{dd}}.
\label{eqn:Dstrength}
\end{align}
Convenient dimensionless parameters can be defined in terms of a reference frequency $\bar \omega $ and the corresponding oscillator length $l=\sqrt{\hbar/(m\bar \omega)}$. Using dimensionless variables $\mathbf{r}' = {\bf r}/l, a' = a/l, a_{\mathrm{dd}}'= a_{\mathrm{dd}}/l$, $t' = t\bar \omega$, $x' = x/l$, $y' = y/l$, $z' = z/l$, $\Omega'=\Omega/\bar \omega$, $\phi' = l^{3/2}\phi$, Eq.~(\ref{eqn:dgpe}) can be rewritten (after dropping the primes from all the variables) as
\begin{subequations}%
\begin{align}\label{gpe3d}
\mathrm{i} \frac{\partial \phi( {\mathbf {r}},{t})}{\partial t} = & \left[-\frac{1}{2}\nabla^2 + V(r) + 4 \pi a N \vert \phi({\mathbf { r}},{ t})\vert ^2 \right. \notag \\
+ & \left. D \int V_{\mathrm{dd}}(\mathbf{r}-\mathbf{r}')\vert \phi( {\mathbf { r}}', t) \vert ^2 d {\mathbf { r}}' \right] \phi({\mathbf { r}},{ t}),
\end{align}
with
\begin{align}\label{vr}
V(r) & = \frac{1}{2}\left({\gamma^2 x^2+\nu^2 y^2}+\lambda^{2} z^2\right),\\
V_{\mathrm{dd}}(\mathbf{r}-\mathbf{r}')& =\frac{1-3\cos^2\theta}{\vert \mathbf{r}-\mathbf{r}' \vert^3}\left(\frac{3\cos^2{\varphi}-1}{2}\right),\label{vdd}
\end{align}
\end{subequations}%
$\gamma= \omega_x/\bar \omega$, $\nu=\omega_y/\bar \omega$, and $\lambda =\omega_z/\bar \omega$. We consider the cylindrically symmetric harmonic trap with $\gamma=\nu$ with $\omega_x=\omega_y=\omega_\rho$ and we use the reference frequency $\bar \omega$ as $\omega_\rho$. From now, we refer only to the dimensionless variables. For our present study, we consider the stationary solutions of Eq.~(\ref{gpe3d}), that is, $\phi(\mathbf r)$.

We perform numerical simulation of the 3D GP Eq.~(\ref{gpe3d}) using the split-step Crank-Nicolson method described in Ref.~\onlinecite{Gammal2006, cpc1, cpc2, cpc3}. The dipolar integral in Eq.~(\ref{gpe3d}), diverges at short distance in coordinate space. However, this can be circumvented by evaluating the integral in momentum space\cite{Goral2002, cpc4, cpc5, cpc6}. The numerical simulations are carried out with $128\times 128\times 128$ grid size, with $\Delta x = \Delta y = \Delta z = 0.2$ (space step) and $\Delta t = 0.003$ (time step).

\section{Calculation of The Shannon information entropy}
\label{sec:3}
The Shannon information entropy in position space, $S_r$, for the density distribution $\rho(\mathbf{r})$ is calculated by
\begin{align}
S_r = -\int \rho(\mathbf{r}) \ln \rho(\mathbf{r}) d\mathbf{r},
\label{eqn:sr}
\end{align} 
where $\rho({\mathbf r} ) = \vert \phi({\mathbf r})\vert^2$ is the one body density
and the corresponding information entropy in momentum space $S_{k}$ is calculated as 
\begin{align}
S_k = -\int n(\mathbf{k}) \ln n(\mathbf{k}) d\mathbf{k}
\label{eqn:sk}
\end{align} 
{where, $n(\mathbf{k}) = \vert\tilde{\phi}(\mathbf{k})\vert^{2}$ is the density distribution in the momentum space, and the momentum space wavefunction, $\tilde{\phi}(\mathbf{k})$, is obtained from the fast Fourier transform of $\phi({\mathbf r})$.} $S_r $ and $S_k$ are calculated by following a similar approach as given in Ref.~\onlinecite{Massen2002}.

We solve numerically Eq.~(\ref{eqn:dgpe}) for two systems: $^{52}$Cr and $^{164}$Dy. The choice of $^{52}$Cr and $^{164}$Dy has a significance for the present study. $^{52}$Cr has a relatively smaller characteristic dipole length $a_{dd}\simeq 16a_0$, while $^{164}$Dy has a larger characteristic dipole length $a_{dd}\simeq 131a_0$. This contrast in the characteristic dipole lengths aids to understand the effect of dipole-dipole interaction strength $D$ (since $D=3N a_{dd}$) on the information entropy. For the present study we fix the inter-atomic scattering $a$, which is equal to $10a_0$. The accuracy of the numerical results is ensured by repeating the calculations with different step sizes and also verified by reproducing the results of Ref.~\onlinecite{Massen2002}. The total entropy $S=S_r+S_k$ as a function of number of bosons $N$ is calculated for several trap geometries and presented in Table~\ref{t1} and Table~\ref{t2}.%

\begin{table*}[htb!]
\caption{Values of $S_r$, $S_k$ and $S$ with lower and upper bounds for three different trap aspect ratios $\lambda=0.5$, $1$ and $2$ of $^{164}$Dy dipolar bosonic system. Here $N$ is the number of bosonic atoms.}
\begin{tabular}{cccrrrrrrcr}
\toprule
\multicolumn{1}{c}{$\lambda$} & \multicolumn{1}{c}{$N$} & \multicolumn{1}{c}{${S_r}_{\mbox{min}}$} & \multicolumn{1}{c}{$S_r$} 
    & \multicolumn{1}{c}{${S_r}_{\mbox{max}}$} & \multicolumn{1}{c}{${S_k}_{\mbox{min}}$} & \multicolumn{1}{c}{$S_k$} 
     & \multicolumn{1}{c}{${S_k}_{\mbox{max}}$} & \multicolumn{1}{c}{${S}_{\mbox{min}}$} & \multicolumn{1}{c}{$S=S_r+S_k$} 
     & \multicolumn{1}{c}{${S}_{\mbox{max}}$} \\ 
\midrule
\multirow{5}{2em}{0.5}&$5\times 10^2$ & $3.635$ & $3.728$ & $3.836$ & $2.598$ & $2.731$ & $2.799$ & $6.434$ & $6.459$ & $6.635$\\
&$10^3$ & $3.708$ & $3.812$ & $3.934$ & $2.501$ & $2.641$ & $2.727$ & $6.434$ & $6.453$ & $6.660$ \\
&$10^4$ & $4.299$ & $4.547$ & $4.794$ & $1.641$ & $1.951$ & $2.135$ & $6.434$ & $6.498$ & $6.928$ \\
&$10^5$ & $5.166$ & $5.681$ & $6.062$ & $0.372$ & $1.008$ & $1.268$ & $6.434$ & $6.690$ & $7.330$ \\
&$10^6$ & $6.148$ & $6.976$ & $7.424$ & $-0.989$ & $-0.014$ & $0.286$ & $6.434$ & $6.962$ & $7.709$ \\
\midrule
\multirow{5}{2em}{1.0} &$5 \times 10^2$ & $4.101$ & $4.187$ & $4.211$ & $2.223$ & $2.293$ & $2.333$ & $6.434$ & $6.480$ & $6.544$ \\
&$10^3$ & $4.291$ & $4.417$ & $4.451$ & $1.984$ & $2.081$ & $2.143$ & $6.434$ & $6.498$ & $6.594$ \\
&$10^4$ & $5.230$ & $5.612$ & $5.695$ & $0.740$ & $1.137$ & $1.204$ & $6.434$ & $6.749$ & $6.899$ \\
&$10^5$ & $6.247$ & $6.929$ & $7.046$ & $-0.612$ & $0.183$ & $0.187$ & $6.434$ & $7.112$ & $7.233$ \\
&$10^6$ & $7.318$ & $8.289$ & $8.422$ & $-1.988$ & $-1.000$ & $-0.883$ & $6.434$ & $7.289$ & $7.538$ \\
\midrule
\multirow{5}{2em}{2.0}&$5 \times 10^2$ & $2.959$ & $3.068$ & $3.159$ & $3.275$ & $3.438$ & $3.475$ & $6.434$ & $6.506$ & $6.597$ \\
&$10^3$ & $3.041$ & $3.165$ & $3.266$ & $3.168$ & $3.335$ & $3.393$ & $6.434$ & $6.499$ & $6.659$ \\
&$10^4$ & $3.661$ & $3.955$ & $4.151$ & $2.283$ & $2.563$ & $2.774$ & $6.434$ & $6.517$ & $6.925$ \\
&$10^5$ & $4.523$ & $5.115$ & $5.422$ & $1.012$ & $1.621$ & $1.911$ & $6.434$ & $6.735$ & $7.333$ \\
&$10^6$ & $5.501$ & $6.418$ & $6.784$ & $-0.350$ & $0.611$ & $0.933$ & $6.434$ & $7.029$ & $7.717$ \\
\bottomrule
\end{tabular}
\label{t1}
\end{table*}

In Table~\ref{t1}, we present the values of $S_r$, $S_k$ and $S$ along with lower and upper bounds for $^{164}$Dy BEC with different trap aspect ratios considering a wide range of number of particles. 
{For $\lambda<1$ the trap is said to be prolate and for $\lambda>1$ the trap is called oblate and spherical for $\lambda=1$.}

The lower and upper bounds in the different entropy measures have been discussed by Gadre and Bendale\cite{Gadre1987}. {We  test the inequalities (\ref{eqn:inequality1})-(\ref{eqn:inequality3}) given in appendix \ref{app:1}, which provides the lower bound as well as upper bound of the total entropy and information entropy in individual spaces.} For a pure spherical trap ($\lambda=1$) with $N=500$, $6.434 \le S_r + S_k \le 6.544$, the actual value of information entropy sum $S_r+S_k$ is $6.48$, which is within $0.71\%$ to the lower bound and $0.97\%$ to the upper bound. The corresponding bounds to information entropy in the individual spaces are $4.101 \le S_r \le 4.211$ (about $2\%$ to the lower bound and $0.56\%$ to the upper bound) and $2.223 \le S_k \le 2.333$ (about $3.14\%$ to the lower bound and $1.71\%$ to the upper bound). For such low $N$ limit in the spherical trap, the total entropy is close to lower bound as the effect of interaction is not {important}. For a larger number of particles, for instance, $N = 10^{5}$, utilizing the inequality relation, we get the total entropy within $13.3\%$ to the lower bound and $3.3\%$ to the upper bound. It demonstrates the effect of interaction which pushes the total entropy towards the upper bound. Similar features are also observed in the information entropy in the individual spaces. 

For $\lambda=0.5$, the trap is prolate, the freedom is restricted and the dipoles can align only along one direction. From Table~\ref{t1}, it is seen that the total entropy and $S_{r}$ are always smaller than those observed in a spherical trap. {As the degrees of freedom are reduced in the trap it leads to a more} ordered state due to dipolar interaction. {On the other hand, the $S_{k}$ values are larger than those in the spherical trap which illustrates that reducing the degree of freedom leads to disorder in momentum space.} For $N=10^{4}$, the total entropy is closer to the lower bound by $0.99\%$ and to the upper bound by $6.2\%$. It is to be noted that at such large particles limit the total effective interaction would be dominating if it were a spherical trap and would have results which are close to the upper bound. However, the anisotropic effect of the prolate trap together with the dipole-dipole interaction make the system more favourable towards the lower bound.

Similarly, for $\lambda = 2$ (oblate trap), we have more ordered states as observed in a prolate trap. For $N=10^5$, the total entropy is again close to the lower bound only by $4.6\%$, whereas to upper bound by $8.1\%$. Thus the net effect of dipolar interaction together with the anisotropic trap is to lead the system into an ordered state. The above investigation of the inequalities and tightness of the upper and lower bounds {clearly} demonstrates this effect.

From Table~\ref{t2}, we also observe that $S_r$ increases with $N$ for all trap geometries while $S_k$ decreases. %
\begin{table}[htb!]
\centering
\caption{The same as in Table~\ref{t1} without lower and upper bounds for $^{52}$Cr and $^{164}$Dy dipolar bosonic system.}
\label{t2}
\begin{tabular}{rrrrrrrrrrr} 
\toprule
\multicolumn{2}{c}{}&\multicolumn{2}{c}{$S_r$} & \multicolumn{2}{c}{$S_k$} & \multicolumn{2}{c}{$S$} \\ 
\cmidrule(r){3-4} \cmidrule(r){5-6} \cmidrule(r){7-8} 
 \multicolumn{1}{c}{$\lambda$} & \multicolumn{1}{c}{$N$} &  \multicolumn{1}{c}{$^{52}$Cr} &  \multicolumn{1}{c}{$^{164}$Dy} 
	&  \multicolumn{1}{c}{$^{52}$Cr} &  \multicolumn{1}{c}{$^{164}$Dy} &  \multicolumn{1}{c}{$^{52}$Cr} &  \multicolumn{1}{c}{$^{164}$Dy}\\ 
\midrule
&$5\times 10^2$    &$3.663$    &$3.728$    &$2.803$    &$2.731$    &$6.465$    &$6.459$ \\   
&$10^3$            &$3.719$    &$3.812$    &$2.740$    &$2.641$    &$6.460$    &$6.453$ \\  
0.5 &$10^4$        &$4.314$    &$4.547$    &$2.153$    &$1.951$    &$6.467$    &$6.498$ \\
&$10^5$            &$5.377$    &$5.681$    &$1.249$    &$1.008$    &$6.626$    &$6.690$ \\
&$10^6$            &$6.645$    &$6.976$    &$0.243$    &$-0.014$   &$6.888$    &$6.962$ \\
\midrule
&$5\times 10^2$    &$4.177$    &$4.187$    &$2.302$    &$2.293$    &$6.479$    &$6.480$ \\ 
&$10^3$            &$4.406$    &$4.417$    &$2.091$    &$2.081$    &$6.497$    &$6.498$ \\
1.0 &$10^4$        &$5.598$    &$5.612$    &$1.146$    &$1.137$    &$6.745$    &$6.749$ \\
&$10^5$            &$6.915$    &$6.929$    &$0.192$    &$0.183$    &$7.107$    &$7.112$ \\
&$10^6$            &$8.275$    &$8.289$    &$-0.982$   &$-1.000$   &$7.292$    &$7.289$ \\
\midrule
&$5\times 10^2$    &$2.991$    &$3.068$    &$3.522$    &$3.438$    &$6.513$    &$6.506$ \\  
&$10^3$            &$3.057$    &$3.165$    &$3.449$    &$3.335$    &$6.507$    &$6.499$ \\      
2.0 &$10^4$        &$3.710$    &$3.955$    &$2.789$    &$2.563$    &$6.499$    &$6.517$ \\    
&$10^5$            &$4.806$    &$5.115$    &$1.851$    &$1.621$    &$6.657$    &$6.735$ \\    
&$10^6$            &$6.086$    &$6.418$    &$0.852$    &$0.611$    &$6.938$    &$7.029$ \\
\bottomrule
\end{tabular}
\end{table}%
Thus, the position space density delocalizes while the momentum space density localizes. For all trap geometries, $S_r$ of $^{164}$Dy is greater than that of $^{52}$Cr. This is in line with the interpretation that larger interactions result in a more delocalized position space density. In $S_k$, the converse is true. Larger interactions lead to a more localized momentum density. It is interesting that the components behave as one would expect, independent of the geometry, while $S$ exhibits deviations when one goes away from the spherical {limit}. Furthermore, for small values of $N$, the geometric effect of the trap dominates over dipole-dipole interaction, as a result, there is only a slight difference in $S$ values between $^{52}$Cr and $^{164}$Dy. However, the dipole-dipole interaction dominates over the geometric effect of the trap for large $N$ values. Thus, the Shannon entropy of $^{52}$Cr and $^{164}$Dy show a considerable difference and this holds for all the three trap geometries.

\begin{figure}[htpb]
\begin{center}
\includegraphics[width=0.99\columnwidth]{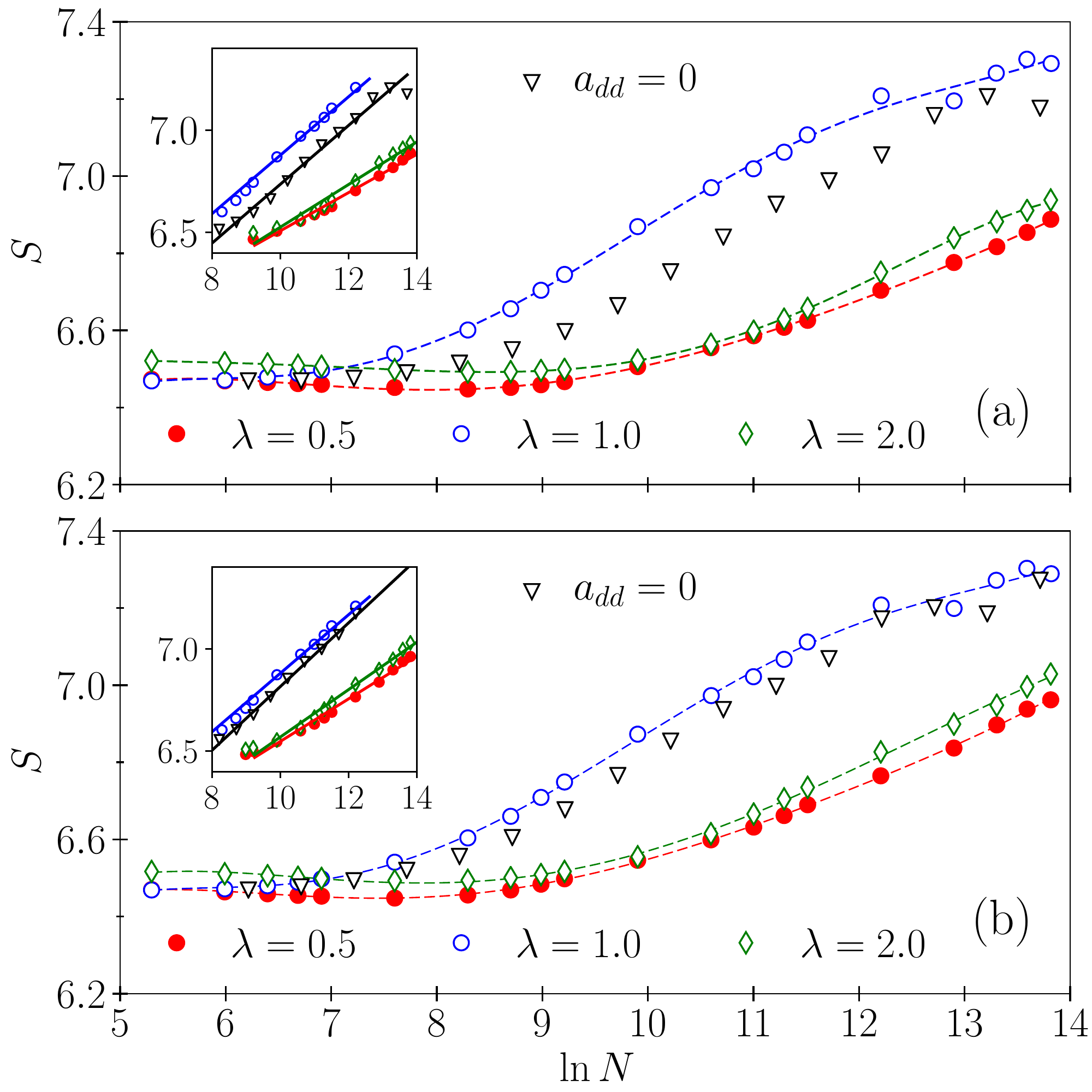}
\end{center}
\caption{The information entropy $S$ versus $\ln N$ for dipolar bosonic systems $^{52}$Cr (a) and $^{164}$Dy (b) for three different trap aspect ratios $\lambda=0.5$ (filled red circle), $\lambda=1$ (empty blue circle) and $\lambda=2$ (green diamond). The dashed lines correspond to the respective fits with Eq.~(\ref{eqn:fitsn}). Black inverted triangles correspond to the case with $a_{dd} = 0$, $a = 10a_0$ and $\lambda = 1$. Insets show the straight line fits of the respective data.}
\label{fig1}
\end{figure}

In Fig.~\ref{fig1}, we plot the Shannon information entropy as a function of the logarithm of the number of dipolar bosonic atoms $N$ for three different values of trap aspect ratio ($\lambda = 0.5$, $\lambda = 1$, and $\lambda = 2$). 
For non-dipolar BEC in a spherical trap, a linear relation between $S$ and $\ln N$ has been prescribed~\cite{Massen2001} for the whole range of particle number. However, in our present calculation, we observe a strong effect of the anisotropy in the entropy measure. So {we do not get a} linear relation for the entire range of particle number. In the insets of Fig.~\ref{fig1}, we show the range of $N$ where a linear relation between $S$ and $\ln N$ holds. On the other hand, one could better fit it with a quintic polynomial. For instance, our numerical results {appear to be well fitted by} 
a quintic polynomial form:
\begin{align}
S=\sum_{j=0}^{5} \alpha_j(\ln N)^j
\label{eqn:fitsn}
\end{align}
where, $\alpha_j$'s are given in Table~\ref{t3}.
\begin{table*}[htpb]
\caption{Values of $\alpha_i$'s in Eq.~(\ref{eqn:fitsn}).}
\begin{tabular}{rrrrrrrrrrrrr} 
\toprule
& \multicolumn{2}{c}{$\alpha_0$} & \multicolumn{2}{c}{$\alpha_1$} & \multicolumn{2}{c}{$\alpha_2$}&\multicolumn{2}{c}{$\alpha_3$} 
       & \multicolumn{2}{c}{$\alpha_4$}&\multicolumn{2}{c}{$\alpha_5$} \\ 
\cmidrule(r){2-3} \cmidrule(r){4-5} \cmidrule(r){6-7}\cmidrule(r){8-9}\cmidrule(r){10-11} \cmidrule(r){12-13}
\multicolumn{1}{c}{$\lambda$} &  \multicolumn{1}{c}{$^{52}$Cr} &  \multicolumn{1}{c}{$^{164}$Dy} &  \multicolumn{1}{c}{$^{52}$Cr} 
      &  \multicolumn{1}{c}{$^{164}$Dy} &  \multicolumn{1}{c}{$^{52}$Cr} &  \multicolumn{1}{c}{$^{164}$Dy} &  \multicolumn{1}{c}{$^{52}$Cr} 
      &  \multicolumn{1}{c}{$^{164}$Dy} &  \multicolumn{1}{c}{$^{52}$Cr} &  \multicolumn{1}{c}{$^{164}$Dy} &  \multicolumn{1}{c}{$^{52}$Cr} 
      &  \multicolumn{1}{c}{$^{164}$Dy} \\ 
\midrule
$0.5$ & $2.71$ & $2.90$ & $2.15$ & $2.14$ & $-0.46$ & $-0.49$ & $0.05$ & $0.05$ & $-0.03$ & $-0.03$ & $0.04$ & $0.05$  \\
$1.0$ & $-1.12$ & $-0.57$ & $4.81$ & $4.48$ & $-1.17$ & $-1.10$ & $0.14$ & $0.13$ & $-0.01$ & $-0.01$ & $0.01$ & $0.01$\\
$2.0$ & $8.61$ & $3.20$ & $-1.44$ & $1.90$ & $0.39$ & $-0.38$ & $-0.05$ & $0.04$ & $0.02$ & $-0.02$ & $-0.07$ & $0.02$ \\
\bottomrule
\end{tabular}
\label{t3}
\end{table*}
The range of values of $N$ used in the fitting procedure obeys both diluteness and quantum degeneracy criteria, which are {needed} for the validity of Eq.~(\ref{eqn:dgpe})\cite{Dalfovo1999}. 
From Eq.~(\ref{eqn:Dstrength}) it is clear that the dipole-dipole interaction 
strength is a linear function of $N$ and so we can write from Eq.~(\ref{eqn:Dstrength}),
\begin{align}
\ln N = \ln(1/3 a_{dd}) + \ln D
\label{eqn:a_dd}
\end{align}
{Given the considered fit of $S$, one expect that $\ln D$ also follows a similar relation with $S$, as $a_{dd}$ is a constant in our computations}. 
{One may note from Fig.~\ref{fig1} that, for the case of $a_{dd} = 0$, the value of $S$ with respect to $\ln N$ for $^{164}$Dy is higher than that of $^{52}$Cr.  This is essentially from the fact that the effective contact interaction strength of $^{164}$Dy is larger than that of $^{52}$Cr.} 

\begin{figure}[htpb]%
\begin{center}
\includegraphics[width=0.50\textwidth]{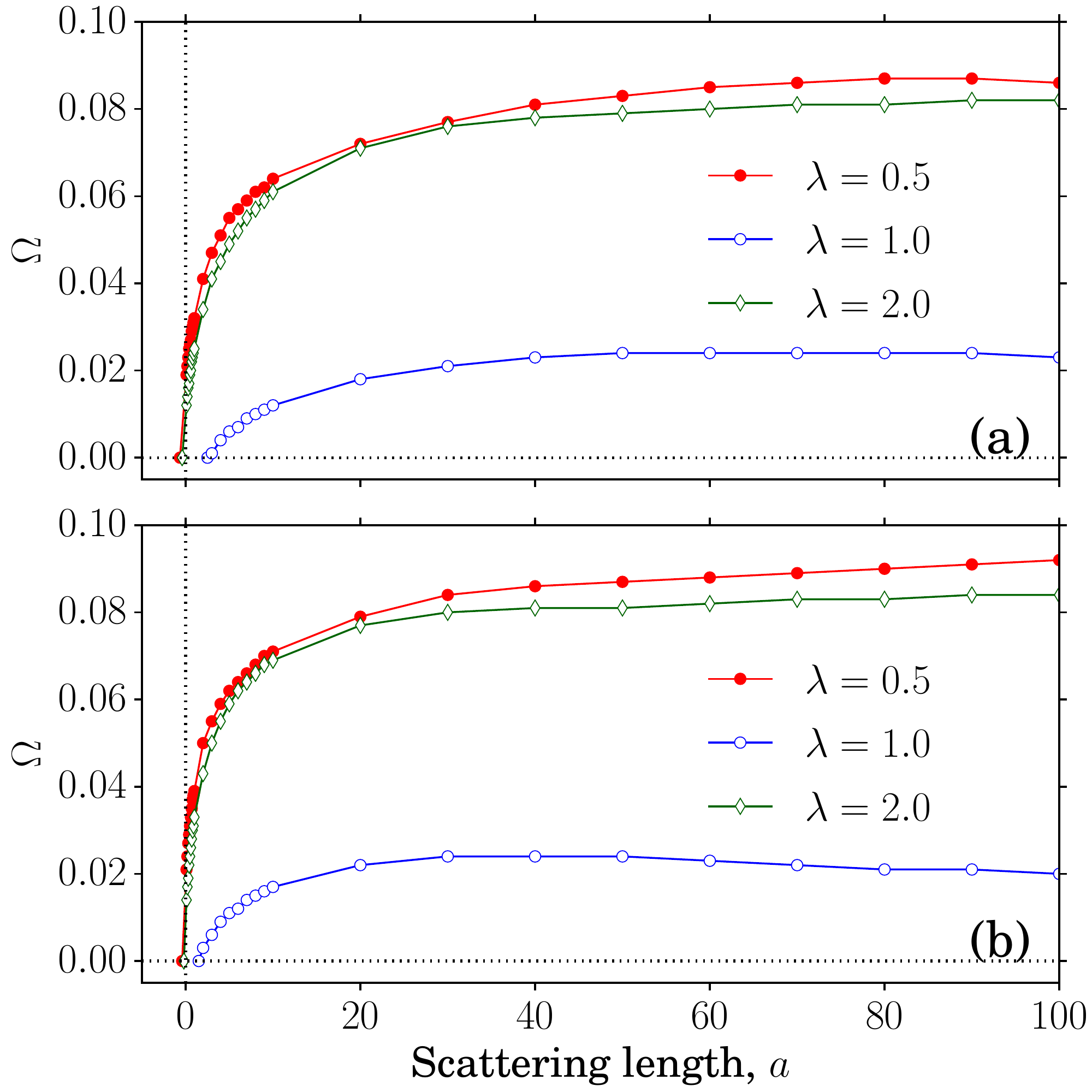}
\end{center}
\caption{The order parameter $\Omega$ versus inter-atomic scattering length ($a$) of dipolar bosonic systems $^{52}$Cr (a) and $^{164}$Dy (b) for three different trap geometries, namely, prolate ($\lambda=0.5$), spherical ($\lambda=1$) and oblate ($\lambda=2$). The number of atoms is fixed as $N=2\times 10^{5}$.}
\label{fig2}
\end{figure}
\begin{figure*}[htpb]%
\begin{center}
\includegraphics[width=0.95\linewidth]{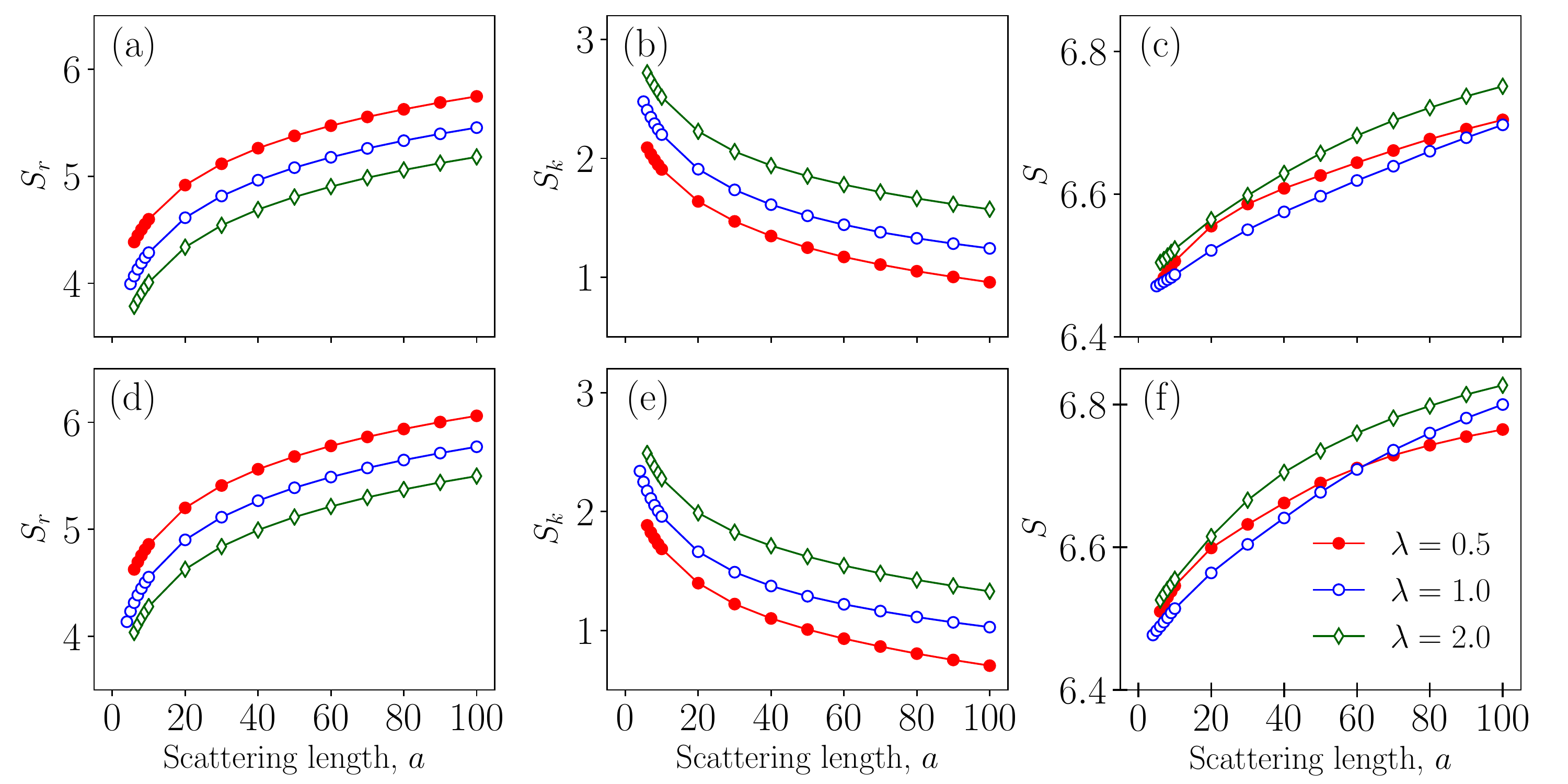}
\end{center}
\caption{The components ($S_r$ and $S_k$) and the total information entropy ($S$) are plotted against the inter-atomic scattering length ($a$) of dipolar BECs of $^{52}$Cr (a)-(c) and $^{164}$Dy (d)-(f) for three different trap geometries for a fixed number of atoms $N=2\times 10^{5}$.}
\label{fig22}
\end{figure*}

Landsberg established that as disorder and entropy are decoupled, 
{and therefore} it is a generic question whether both entropy and order can increase together~\cite{Landsberg1984}. Landsberg defined the order parameter $\Omega$ as
\begin{align}
\Omega = 1-\frac{S}{S_{\mbox{max}}}
\label{eqn:omega}
\end{align}
where $S$ is the total information entropy of the system and $S_{\mbox{max}}$ the maximum entropy accessible to the system. $\Omega=0$ implies system is at maximum accessible entropy as for this case $S=S_{\mbox{max}}$ and from the information-theoretical point of view, the system is in a completely disordered state and random. $\Omega=1$ implies the system is at zero entropy. However, for a realistic and natural system, $\Omega$ lies between $0$ and $1$. This is indeed required to satisfy the Heisenberg uncertainty relation together with EUR. In the present context, it is seen how the increase in interaction strength (that is, the scattering length $a$) gradually leads the system from a disordered to ordered state. For such a complex system where the trap geometry plays a 
{crucial} role in the disorder to order transition, we further study the variation of $\Omega$ with $a$ for fixed number of particles.

In Fig.~\ref{fig2}, we plotted $\Omega$ versus the inter-atomic scattering length $a$. By varying $a$, the strength of repulsive interaction can be tuned. For the present investigation, we consider $^{52}$Cr and $^{164}$Dy dipolar BECs with $N=2\times10^5$ atoms. This value of $N$ is within the range such that it obeys both diluteness and quantum degeneracy criteria. It may be noted from Fig.~\ref{fig2} that $\Omega$ for a spherical trap is considerably lower than the anisotropic traps for the whole range of scattering length. As we have mentioned 
earlier, that is due to the excess of freedom available in the spherical trap.  The dipoles are initially in a disordered state for small values of $a$, however, with the increase in scattering length $\Omega$ gradually increases and reaches its saturation. Thus for the spherical trap, the disorder to order transition is mainly due to the effect of an increase in the scattering length. The situation is more complicated when we move from spherical to asymmetric traps. For both prolate and oblate traps, as the system is already in a state which is more ordered compared to the corresponding state for the spherical trap (due to the reduction of degrees of freedom), the value of $\Omega$ is higher than that of the spherical trap. The steep increase in $\Omega$ for a slight change in scattering length leads to a sharp change in disorder to order state and the saturation value of $\Omega$ is higher than that of a spherical trap. This observation is in good agreement with the previous analysis of Table~\ref{t1}, where disorder to order transition is manifested through the lower and upper bounds of the entropy inequalities. Another interesting point is to study the threshold value of the scattering length for which the order parameter $\Omega$ tends to zero. 
\begin{table}[htpb]
\caption{The threshold values of atomic scattering $a_{thres}$ in Fig.~\ref{fig2}}
  \begin{tabular}{crr}
    \hline
    \multirow{2}{*}{$\lambda$} &
      \multicolumn{2}{c}{$a_{thres}$} \\
\cmidrule(r){2-3}
    & $^{52}$Cr & $^{164}$Dy  \\
    \hline
    $0.5$ & $-0.65$ & $-0.4$ \\
    $1.0$ & $2.50$ & $1.5$\\
    $2.0$ & $-0.40$ & $-0.2$ \\
    \hline
  \end{tabular}
\label{t4}
\end{table}
According to inequality criteria, $S$ is always less than $S_{\text{max}}$, {and one cannot reach the $\Omega = 0$ value. However, we define the 
threshold} scattering length for which $\Omega$ becomes very close to zero, which quantifies the maximum possible disorder in the system. In Table~\ref{t4} the corresponding threshold values for all the considered 
trap geometries and for both $^{52}$Cr and $^{164}$Dy are presented. 
Following our previous discussion, it is expected that $a_{thres}$ {is larger} for spherical trap than that of asymmetric traps. As the system is already in an ordered state both for the prolate and oblate trap, even when the scattering length is zero, one has to {make $a_{thres}$} negative to reach the disordered state. We observe a similar behavior in $a_{thres}$ for$^{52}$Cr and $^{164}$Dy BECs.

We also plot the individual components ($S_r$ and $S_k$) and the total information entropy ($S$) with respect to the scattering length in Fig.~\ref{fig22}. For all trap geometries, the position component $S_{r}$ increases with the increase of the repulsive interaction strength $a$, while the momentum component $S_{k}$ decreases with $a$. Larger repulsive interaction results in delocalization in position space density and localization in momentum space density. The entropy sum $S$ increases with $a$.

\begin{figure}[htpb]
\begin{center}
\includegraphics[width=0.99\linewidth]{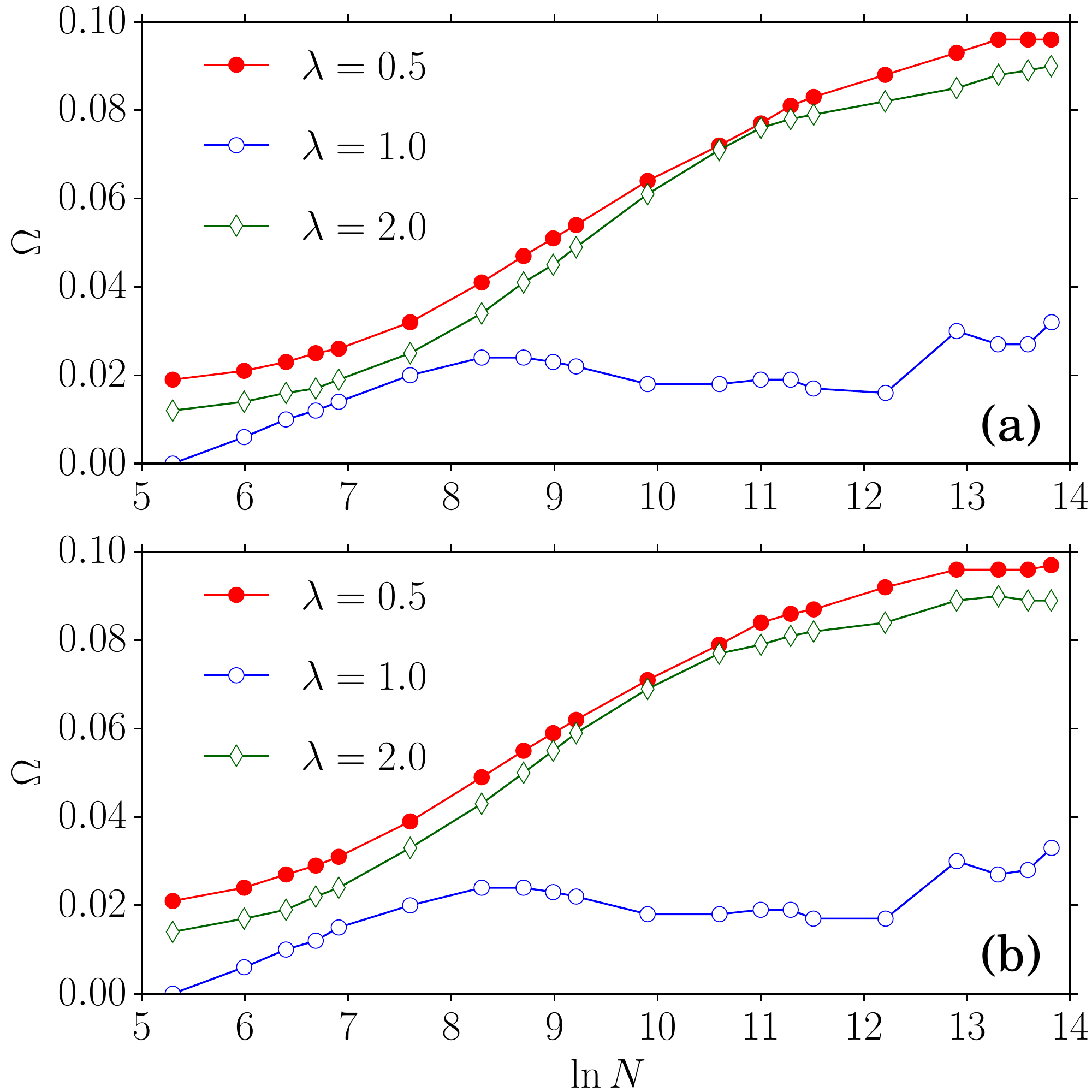}
\end{center}
\caption{The order parameter $\Omega$ as a function of $\ln N$ for $^{52}$Cr (a) and $^{164}$Dy (b) atoms for three different trap aspect ratios $\lambda=0.5$ , $\lambda=1$ and $\lambda=2$.}
\label{fig3}
\end{figure}

Next, in Fig.~\ref{fig3}, we plot the order parameter $\Omega$ as a function of $\ln N$ for several trap geometries and for $^{52}$Cr and $^{164}$Dy atoms respectively. It is seen that $\Omega$ is an increasing function of $N$ for both systems. %
{Fig.~\ref{fig3}} indicates that as particles are added, 
the system becomes more ordered and leads to saturation. For a spherical trap, the dipoles are in a more disordered state than in the anisotropic trap due to the availability {of extra degrees of freedom.} Even for a large number of particles, we do not observe saturation in $\Omega$ and the value of it is very small. Whereas for both prolate and oblate trap due to the 
restriction of the motion of the dipoles, {the system is initially in a more ordered state.} Thus in Fig.~\ref{fig3} we observe that even for small $N$, $\Omega$ is larger than that of a spherical trap. With an increase in particle number, it leads to a quick increase in $\Omega$. 
{This result is an agreement with} our earlier findings that dipolar interaction in the anisotropic trap basically pushes the system towards more ordered states.

{This result is in agreement} with the earlier observation made by Landsberg and Shiner\cite{Landsberg1998}, where it is  shown  that $\Omega$ is small for a small number of electrons and it increases, as one pumps more electrons into the system which fills up the energy levels. 
\begin{figure}[htpb]
\begin{center}
\includegraphics[width=0.99\linewidth]{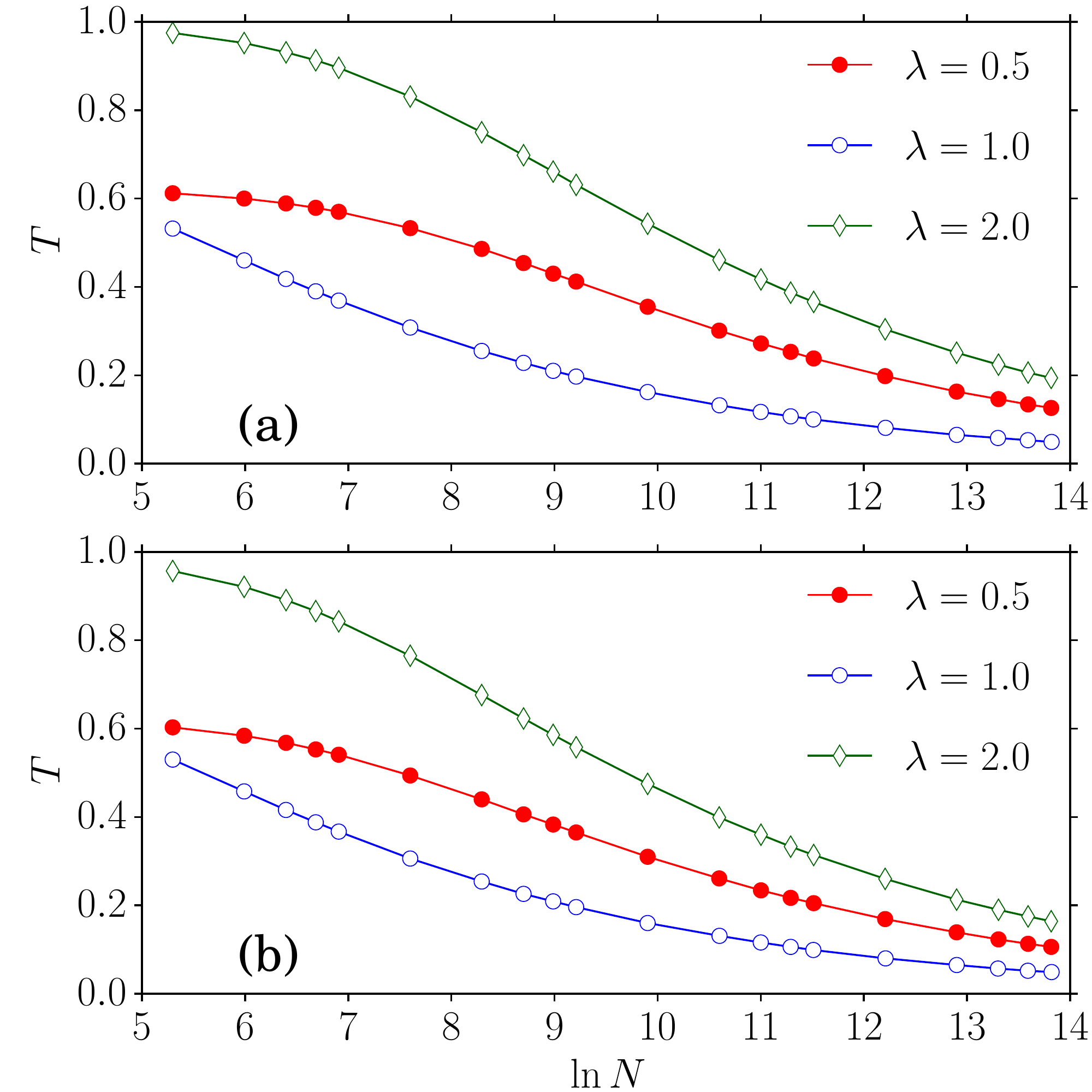}
\end{center}
\caption{Total kinetic energy $T$ of the system as a function of $\ln N$ for dipolar BECs of (a) $^{52}$Cr atoms and (b) $^{164}$Dy atoms for three different trap aspect ratios $\lambda=0.5$ , $\lambda=1$ and $\lambda=2$.}
\label{fig4}
\end{figure}
In Fig.~\ref{fig4}, we plot the total kinetic energy $T$ as a function of $\ln N$. 

In earlier calculations for nuclei and atomic clusters, the relation $T\simeq CN$ is maintained, where $C$ is a constant\cite{Massen2001, Massen2002}; the total kinetic energy per particle is approximately constant. However, in atomic BECs the interaction of atoms are different from that of nuclei and atomic clusters and so the kinetic energy does not have a linear relation with $N$.

In atomic physics, there is already a connection of $S_r$ and $S_{k}$ with the total kinetic energy through some rigorous inequalities as given in the Appendix~\ref{app:1}. 
\begin{table}[htpb]
\caption{Values of $\beta_i$'s of Eq.~(\ref{eqn:fitsT})}
\begin{tabular}{rrrrrrrrr} 
\toprule
& \multicolumn{2}{c}{$\beta_0$} & \multicolumn{2}{c}{$\beta_1$} & \multicolumn{2}{c}{$\beta_2$}&\multicolumn{2}{c}{$\beta_3$}\\
\cmidrule(r){2-3} \cmidrule(r){4-5} \cmidrule(r){6-7}\cmidrule(r){8-9}
 \multicolumn{1}{c}{$\lambda$} &  \multicolumn{1}{c}{$^{52}$Cr} &  \multicolumn{1}{c}{$^{164}$Dy} &  \multicolumn{1}{c}{$^{52}$Cr} &  \multicolumn{1}{c}{$^{164}$Dy} &  \multicolumn{1}{c}{$^{52}$Cr} 
      &  \multicolumn{1}{c}{$^{164}$Dy} &  \multicolumn{1}{c}{$^{52}$Cr} &  \multicolumn{1}{c}{$^{164}$Dy} \\ 
\midrule
$0.5$ & $7.32$ & $7.39$ & $-4.30$ & $-4.96$ & $6.78$ & $8.66$ & $-3.30$ &  $-4.92$ \\
$1.0$ & $7.59$ & $7.59$ & $-6.12$ & $-6.08$ & $10.50$ & $10.40$ & $-5.51$ & $-5.41$ \\
$2.0$ & $7.51$ & $7.53$ & $-3.70$ & $-3.80$ &  $4.37$  & $4.53$ & $-1.66$ & $-1.74$ \\
\bottomrule
\end{tabular}
\label{t5}
\end{table}
{Similar kinds of connections between the Shannon entropy and correlation terms in the kinetic energy functional has been reported in the context of uniform gas of interacting electrons\cite{Ghiringhelli2010}.}
\begin{figure}[htpb]%
\begin{center}
\includegraphics[width=0.99\linewidth]{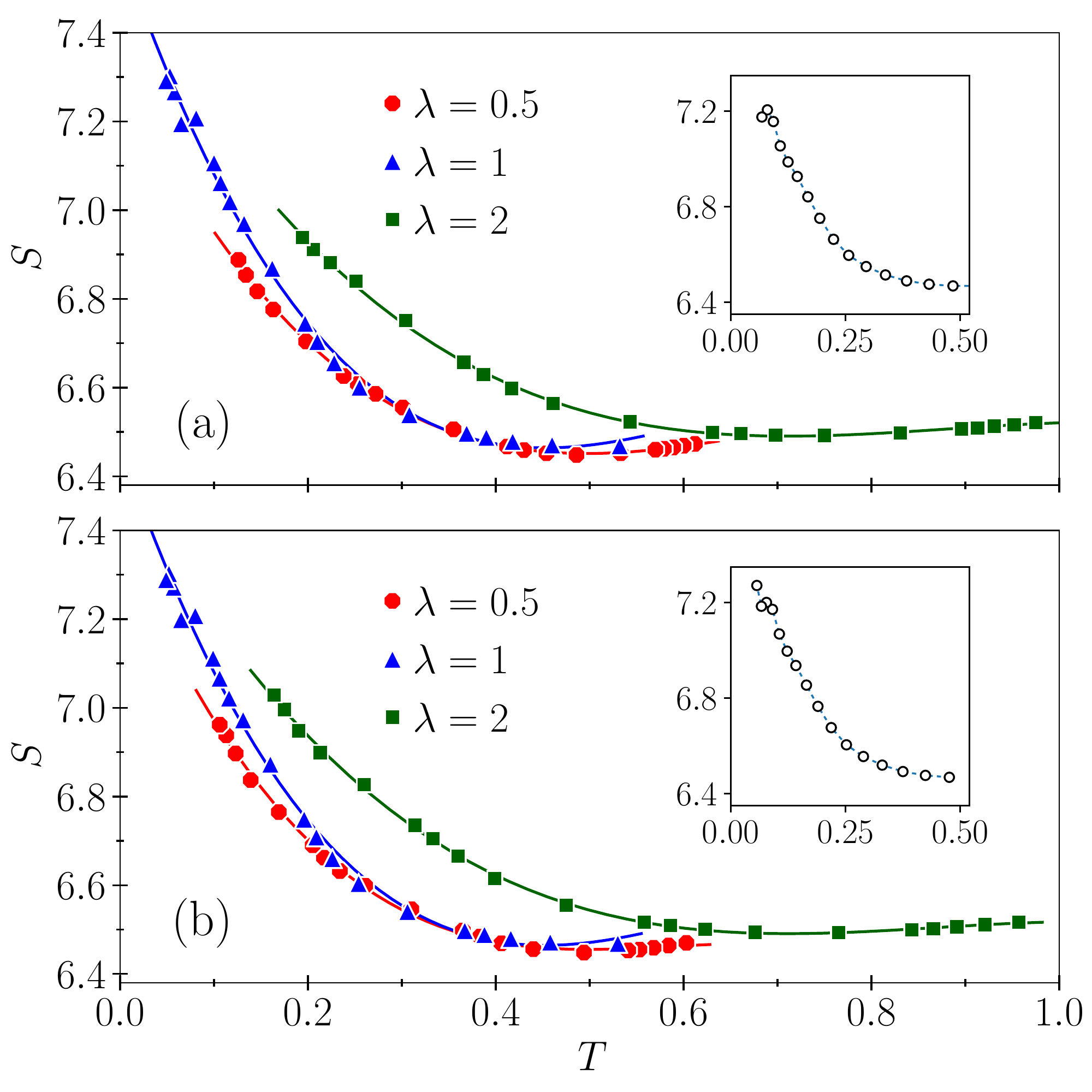}
\end{center}
\caption{The information entropy $S$ versus the total kinetic energy $T$ for dipolar BECs of (a) $^{52}$Cr and (b) $^{164}$Dy atoms for three different trap aspect ratios $\lambda=0.5$ (red circles), $\lambda=1$ (blue triangles) and $\lambda = 2$ (green squares). The solid lines correspond to the respective curve fits with cubic polynomial given in Eq.~(\ref{eqn:fitsT}). The insets in (a) and (b) show the plot of $S$ versus $T$ for the case $^{52}$Cr and $^{164}$Dy BECs, respectively, with $a_{dd} = 0$ and $\lambda = 1$ (spherical trap).}
\label{fig5}
\end{figure}
In the present manuscript, we study as well the link between $S$ and $T$. 
In Fig.~\ref{fig5}, we plot the Shannon information entropy ($S$) versus the total kinetic energy of the system $T$ for three different values of trap aspect ratio ($\lambda = 0.5$, $\lambda = 1$, and $\lambda = 2$). We also plot the Shannon information entropy of the same system in the spherical trap with $a_{dd} = 0$ (see the insets in Fig.~\ref{fig5}). {For smaller values of the kinetic energy, a linear relation between $S$ and $T$ is observed. Our numerical appear to be well fitted by a cubic polynomial form}
\begin{align}
S=\sum_{j=0}^{3} \beta_i(T)^j
\label{eqn:fitsT}
\end{align}
where, $\beta_j$'s are given in Table~\ref{t5}. As pointed out earlier, in nuclei and atomic cluster, the linear relationship between $T$ and $N$ makes $S(N)$ and $S(T)$ to share a similar relationship. However, we did not find such a linear relation between $T$ and $N$ in BECs, and so this straightforward transformation between $S(N)$ and $S(T)$ does not hold for the case of BECs. {The $a_{dd} = 0$ curves, shown in the insets of Figs.~\ref{fig5}(a) and \ref{fig5}(b), also follow a similar cubic form. However, due to the absence of dipolar interaction the value of $S$ with respect to $T$ is smaller than that with $a_{dd} \neq 0$ in the same spherical trap.}

\begin{figure}[htpb]%
\begin{center}
\includegraphics[width=1.0\columnwidth]{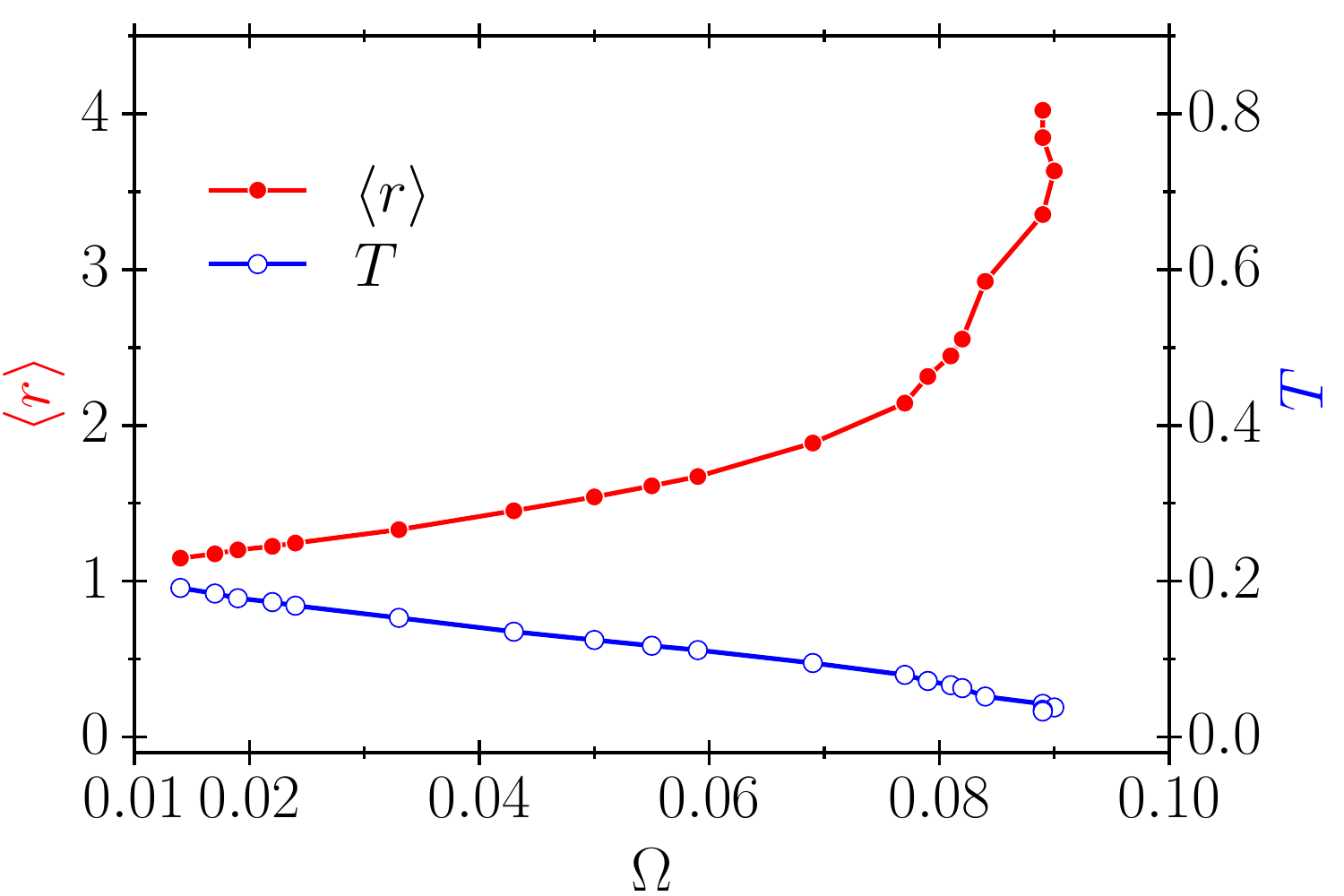}
\end{center}
\caption{The order parameter $\Omega$ versus rms size of the condensate $\langle r \rangle$ and the total kinetic energy $T$ of the system for $^{164}$Dy atoms in an oblate trap. The inter-atomic scattering length is fixed as $a=10a_{0}$.}
\label{fig6}
\end{figure}

In order to get physical insight about the Landsberg order parameter, we plotted in Fig.~\ref{fig6}, 
$\Omega$ versus the rms size of the condensate $\langle r \rangle$ and the total kinetic energy $T$ of the system for $^{164}$Dy atoms in an oblate trap. We observe that as we add more atoms into the system, the size of the system increases and tends towards ordered state which can be inferred from the increasing values of $\langle r \rangle$ and $\Omega$ respectively. A similar kind of trend has been reported for fermions by Landsberg and Shiner\cite{Landsberg1998}. However, the total kinetic energy $T$ decreases with increase in $\Omega$. From this behavior, one could think of kinetic energy as a randomizing factor in a correlated quantum - many body systems.

\section{Summary and Conclusion}
\label{sec:4}

{We studied the Shannon information entropies of one-body density in position and momentum space $S_r$, $S_k$ and their sum $S$ for a dipolar Bose-Einstein condensate (BEC).} We report the results of  both weak and strong dipole-dipole interaction strengths and for different trap geometries. The main motivation of the present manuscript is to study how the effective interaction is affected by the trap geometry and controls the order-disorder transition.   We compare our results with both the non-dipolar BEC and a spherical trap. We observe that the universal trend in the calculation of $S$ and its dependence with $N$ for different fermionic and bosonic systems does not hold for the anisotropic interactions of dipolar BEC. {Although the BBM inequality strictly holds, the trap geometry strongly influences the value of $S$.} The Shannon information entropy $S$ and $\ln N$ {appear to be well described by} a quintic polynomial relation whose parameters strongly depends on trap aspect ratio and also on the dipole-dipole interaction strength. The calculation of the order parameter, its dependence on the inter-atomic scattering length and the number of particles have been studied. We showed  how the anisotropic trap reduces the degrees of freedom of the dipoles and pushes the system into a more ordered state even for very small scattering length. The corresponding non-dipolar BEC in a spherical trap still shows the disordered phase. We observed that adding more particles to the system leads to a more ordered state where the system becomes highly correlated. For the dipolar bosonic system, the Landsberg order parameter becomes vanishing even for a finite number of atoms when the atomic scattering is tuned below the threshold $a_{thres}$. We also make a link between $S$ and the total kinetic energy $T$, which is different from earlier observation for fermionic systems, the dipole-dipole interaction strength and trap aspect ratio play a significant role. However, the relation between $T$ and the order parameter $\Omega$ clearly shows that $T$ can also be taken as a randomizing factor of the highly correlated quantum many-body system. The study of dynamics of Shannon information entropy is also an interesting measure, which is connected with statistical relaxation and eigenstate thermalization hypothesis. {We think that this investigation, requiring the full time-dependent solution of the Gross-Pitaevskii equation, is an interesting subject of future work.}

\begin{acknowledgments}
AT acknowledges discussions with G. Gori and T. Macr\`{i}.  TS acknowledges financial support from University Grants Commission, India in the form UGC-RFSMS fellowship. BC would like to acknowledge financial support from DST major project SR/S2/CMP/0126/2012 and ICTP, Trieste. The work of PM forms a part of Science \& Engineering Research Board (SERB), Department of Science \& Technology (DST), Govt. of India sponsored research project (No. EMR/2014/000644).
\end{acknowledgments}

\appendix

\section{Connection between $S_r$, $S_k$ with the total kinetic energy $T$}
\label{app:1}
Gadre and Bendale\cite{Gadre1987} 
established a connection between $S_r$, $S_k$ with the total kinetic energy $T$ and mean square radius of the system which has been derived using EUR.
\begin{align}
& {S_r}_{\mbox{min}} \leqslant S_r \leqslant {S_r}_{\mbox{max}} , \label{eqn:inequality1}\\
& {S_k}_{\mbox{min}} \leqslant S_k \leqslant {S_k}_{\mbox{max}} ,\label{eqn:inequality2}\\
& {S}_{\mbox{min}} \leqslant S \leqslant {S}_{\mbox{max}} .
\label{eqn:inequality3}
\end{align}
For density distribution normalized to unity, the above lower and upper limits took the form
\begin{subequations}\label{eqn:lowupbound}
\begin{align}
{S_r}_{\mbox{min}} &= \frac{3}{2}(1+ \ln \pi) - \frac{3}{2} \ln \left( \frac{4}{3} T \right), \\ 
{S_r}_{\mbox{max}} &= \frac{3}{2}(1+ \ln \pi) + \frac{3}{2} \ln \left( \frac{2}{3} \langle r^{2} \rangle \right), \\ 
{S_k}_{\mbox{min}} &= \frac{3}{2}(1+ \ln \pi) - \frac{3}{2} \ln \left( \frac{2}{3} \langle r^{2} \rangle \right) , \\ 
{S_k}_{\mbox{max}} &= \frac{3}{2}(1+ \ln \pi) + \frac{3}{2} \ln \left( \frac{4}{3} T \right), \\ 
{S}_{\mbox{min}} &= 3(1+ \ln \pi), \\ 
{S}_{\mbox{max}} &= 3(1+ \ln \pi) + \frac{3}{2} \ln \left( \frac{8}{9} \langle r^{2} \rangle T \right).
\end{align}
\end{subequations}
Massen and Panos\cite{Massen2001} presented the values of the lower and upper bound of $S$ as in Eqs.~(\ref{eqn:lowupbound}). Similar values of the lower and upper bound of $S$ had been calculated\cite{Massen2002} for BECs of $^{87}$Rb and $^{133}$Cs. In the present work, we calculate numerically the values in Eq.~(\ref{eqn:lowupbound}) for $^{52}$Cr and $^{164}$Dy condensates and the results are presented in Tables~\ref{t1} and \ref{t2}.

\end{document}